\documentclass[%
 reprint,
 amsmath,amssymb,
 aps,
]{revtex4-1}

\usepackage{graphicx}
\usepackage{dcolumn}
\usepackage{bm}

\begin{document}

\preprint{APS/123-QED}

\title{Valley depolarization in monolayer WSe$_2$}

\author{Tengfei Yan}
\author{Xiaofen Qiao}%
\author{Pingheng Tan}%
\author{Xinhui Zhang}%
 \email{xinhuiz@semi.ac.cn}
\affiliation{%
State Key Laboratory of Superlattices and Microstructures, Institute of Semiconductors, Chinese Academy of
Sciences, P.O. Box 912, Beijing 100083, People¡¯s Republic of China
}%


\date{\today}

\begin{abstract}
We have systematically examined the circular polarization of monolayer WSe$_2$ at different temperature, excitation energy and exciton density. The valley depolarization in WSe$_2$ is experimentally confirmed to be governed by the intervalley electron-hole exchange interaction. More importantly, a non-monotonic dependence of valley circular polarization on the excitation power density has been observed, providing the experimental evidence for the non-monotonic dependence of exciton intervalley scattering rate on the excited exciton density. The physical origination of our experimental observations has been proposed, which is in analogy to the D$^\prime$yakonov-Perel$^\prime$ mechanism that is operative in conventional GaAs quantum well systems. Our experimental results are fundamentally important for well understanding the valley psudospin relaxation in atomically thin transition metal dichalcogenides.

\end{abstract}

\maketitle



Valley pseudospin, which labels the degenerate energy extrema in momentum space, draws increasing attention with the investigation on two-dimensional (2D) materials as a potential controllable degree of freedom in future electronics. The concept of valleytronics, to analogize spintronics, has been proposed to make use of this degree. Atomically thin transition metal dichalcogenides (TMDCs) is a kind of newly developed 2D materials, with finite direct band gap at the edges of the Brillouin zone at $K_+$ and $K_-$ in monolayer flakes\cite{PhysRevB.89.201302,PhysRevLett.105.136805,PhysRevLett.108.196802,PhysRevB.86.115409}. Monolayer TMDCs exhibit a locking between the light circular dichroism and valley pseudospin due to the broken inversion symmetry and large spin-orbit coupling, making them perfect platforms to investigate valleytronics\cite{PhysRevB.77.235406,xiao2007valley}. Previous works have demonstrated the injection and detection of valley polarization in monolayer TMDCs using photoluminescence (PL) technique\cite{cao2012valley,zeng2012valley,mak2012control,kioseoglou2012valley,PhysRevB.86.081301,PhysRevB.90.075413}. Valley polarization relaxation has also been investigated by the time-resolved photoluminescence(TRPL) technique\cite{PhysRevLett.112.047401,PhysRevB.90.075413}, transient reflectance spectrum\cite{doi:10.1021/nl403742j,doi:10.1021/nn405419h,C4NR03607G} and transient Kerr rotation technique\cite{PhysRevB.90.161302,plechinger2014time}. The reported exciton valley polarization relaxation is surprisingly fast, which is in the picosecond range. And the relaxation process has been shown to be sensitive to temperature, excitation photon energy and sample treatment method.

The previous studies have excluded the D$^\prime$yakonov-Perel$^\prime$ (DP) and Elliott-Yafet (EY) mechanisms to be responsible for the efficient spin relaxation in monolayer MoS$_2$, as the out-of-plane component of exciton spin caused by these mechanisms has been calculated to be in the order of nanoseconds\cite{PhysRevB.89.201302,Wang20141336}, which is much longer than the exciton lifetime. Recently, a Maialle-Silva-Sham (MSS) mechanism caused by the electron-hole (e-h) exchange interaction has been suggested to dominate the spin relaxation in monolayer MoS$_2$ and WSe$_2$\cite{PhysRevB.89.205303,PhysRevB.89.201302,PhysRevB.90.161302}. In this work, we report a systematic circular dichroism resolved photoluminescence study on monolayer WSe$_2$ at different temperatures with widely tuned laser excitation wavelength and power intensity. Our experimental results confirm the important role of the exciton exchange interaction on the valley polarization relaxation. Though the measured valley polarization was observed to decrease with increasing excitation intensity at certain excited exciton density range, as previously reported\cite{PhysRevLett.112.047401}, by varying the exciton density up to four orders of magnitude, a surprising non-monotonic dependence of valley circular polarization on the excitation power density was observed. The non-monotonic dependence suggests the same scenario here as that of the DP-governed spin relaxation in conventional GaAs quantum well structures. The observed non-monotonic dependence of valley polarization relaxation has been proposed to originate from a close analogue of the DP mechanism, in which the exciton intervalley scattering rate is a non-monotonic function of exciton density with a minimum corresponding to the crossover from the non-degenerate regime to the degenerate one by increasing exciton density.


Since the degenerated energy level of A-exciton in $K_+$ and $K_-$ valley corresponds to the opposite spin originating from the broken inversion symmetry, spin flips of electrons and holes are expected to be  essential for the intervalley scattering. The intravalley scattering is energetically forbidden due to the large spin split at K point for TMDC monolayers. It is known that the degree of circular polarization is estimated based on the formula\cite{mak2012control}:
\begin{equation}
 P_c=P_0/(1+2\tau_0/\tau_v)
\end{equation}
in which $P_0$ is the initial polarization degree, $\tau_0$ is the exciton lifetime, so the PL circular polarization is expected to increase monotonically with the intervalley scattering time $\tau_v$. Thus the degree of PL circular polarization is associated with the issue of the intervalley scattering time.

The WSe$_2$ flakes are fabricated by mechanical exfoliation with adhesive tape from a bulk crystal (2D semiconductors Inc.) onto $SiO_2/Si$ substrates. Monolayer WSe$_2$ flakes were identified by optical contrast under a microscope first and then confirmed via Raman and PL measurements as presented in our previous work.\cite{:/content/aip/journal/apl/105/10/10.1063/1.4895471} The circular polarization resolved PL measurements was carried out by a microscopic confocal PL setup\cite{zeng2012valley}. PL response, collimated by a 50$\times$ objective lens, passes through a quarter wave plate and a beam-displacer to be circularly separated out. Excited by a $\sigma^+$ circularly polarized laser beam, the degree of PL circular polarization is determined by $P_c=(I_+-I_-)/(I_++I_-)$, where $I_+$ and $I_-$ correspond to the PL intensity of $\sigma^+$ and $\sigma^-$ component, respectively. A supercontinuum white light source (Fianium Ltd., model: SC450-2) and femtosecond ultrashort pulsed laser (Coherent Inc., model: Chameleon) are used respectively for the purpose to vary the excitation peak density in a wide range. For the temperature dependent measurement, the sample was mounted in an Oxford nitrogen-flow microscope cryostat (Microstat HiRes II).

At temperatures below 180K, two individual peaks are easily distinguished in the steady-state PL response of monolayer WSe$_2$ as shown in Fig. 1, in which the details of typical PL response excited with the pumping energy of 1.77eV at 100K are displayed. The higher energy peak is attributed to the A exciton recombination. While the lower energy PL peak, which has been studied a lot, is related to the trion recombination\cite{jones2013optical,PhysRevB.90.075413,:/content/aip/journal/apl/105/10/10.1063/1.4895471}.
The time-resolved circularly polarized PL response of both the exciton and trion measured at 100K, as shown in Fig. 2, reveals the exciton valley polarization relaxation time to be approximately 24ps. Considering that $\tau_0=36ps$ estimated by TRPL measurement at 100K, $P_c$ is estimated to be 25\% based on equation (1), which is in consistent with the polarization degree determined by our steady-state PL result shown in Fig. 1.

\begin{figure}
\centering
  \includegraphics[width=8cm]{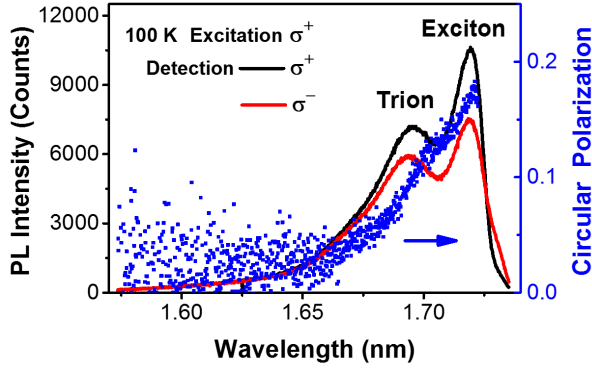}
  \caption{PL response and circular polarization degree of monolayer WSe$_2$ excited with pumping energy of 1.77eV at 100K. The circular polarization degree is plotted in blue dots.}
\end{figure}

\begin{figure}
\centering
  \includegraphics[width=8cm]{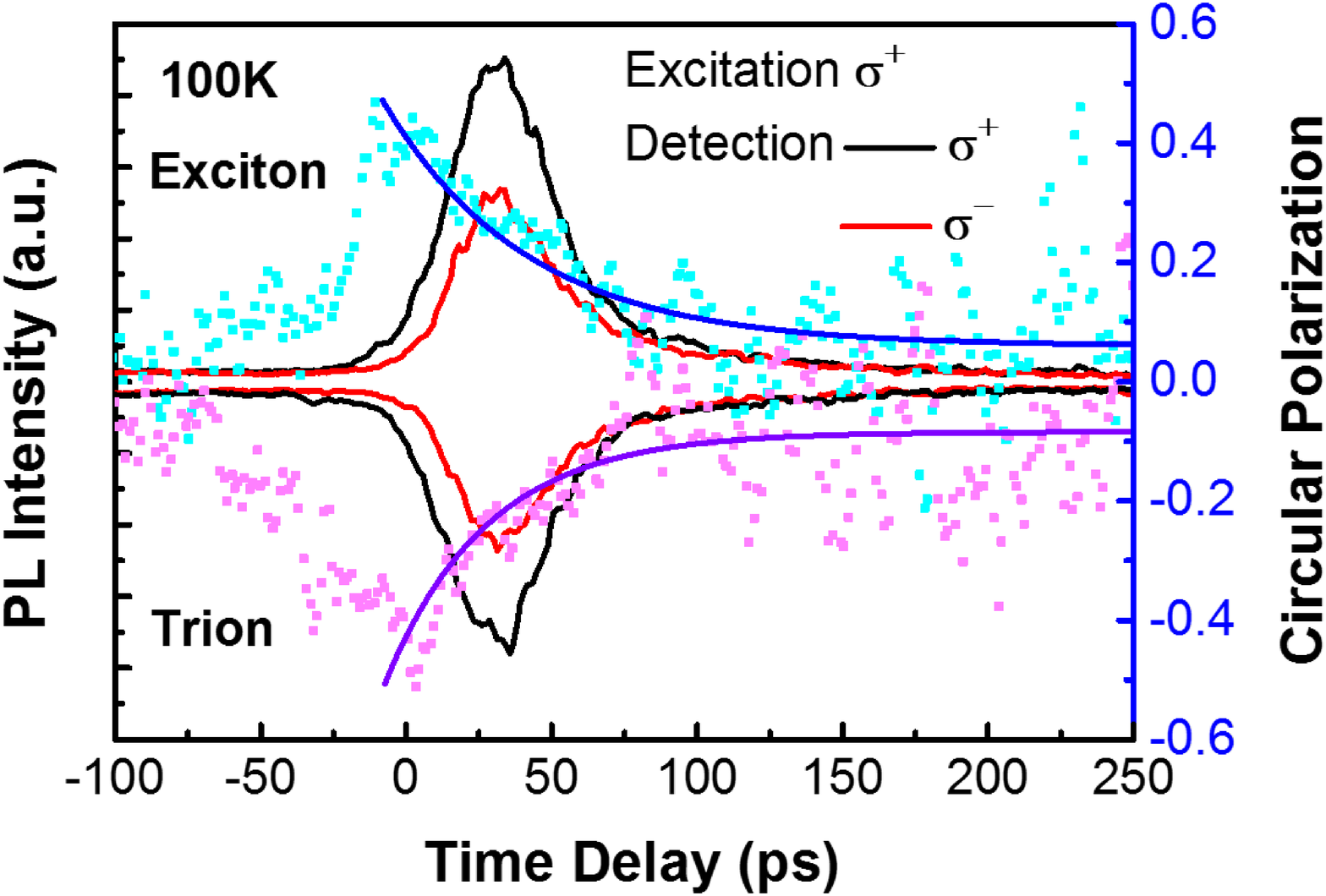}
  \caption{The time-resolved circularly-polarized PL response of both exciton and trion at 100K with pumping of 1.68eV. The valley polarization lifetime is obtained by fitting the exponential decay function as plotted with blue and violet solid lines.}
\end{figure}

Previous studies have suggested that the valley depolarization is governed by the Maialle-Silva-Sham (MSS) mechanism caused by electron-hole(e-h) exchange interaction\cite{PhysRevB.47.15776,jones2013optical,PhysRevB.90.161302,PhysRevB.89.205303,PhysRevB.89.201302}. As has been theoretically pointed out\cite{PhysRevB.89.205303}, the long-range e-h exchange
interaction acting as a momentum-dependent effective magnetic field $\Omega(P)$, where $P$ is the center-of-mass momentum of A exciton. Similar to the DP mechanism, the spin of excitons with different center-of-mass momentums precess around the effective magnetic field with different frequencies, leading to a free-induction decay as that of spin relaxation induced by the inhomogeneous broadening caused by randomly-orientated effective spin-orbit coupling field. Similarly, in analogy to the DP mechanism, the exciton momentum scattering suppresses the inhomogeneous broadening in the strong scattering regime, following the relationship of \cite{PhysRevB.89.205303}:
\begin{equation}
\tau^{-1}_{s(v)}=\langle\Omega^2(P)\rangle\tau^*_p,
\end{equation}
where $\langle\dots\rangle$ denotes the ensemble average and $\tau^*_p$ represents the time of exciton scattering with momentum relaxation. The precession frequency due to the momentum-dependent effective magnetic field resulting from the long-range exchange interaction between the two exciton spin states is written as\cite{PhysRevB.89.205303}:
\begin{equation}
\omega(P)\approx\sqrt5C\alpha(1)|P|/\hbar,
\end{equation}
where C and $\alpha(1)$ are material related parameters.

\begin{figure}
\centering
  \includegraphics[width=8cm]{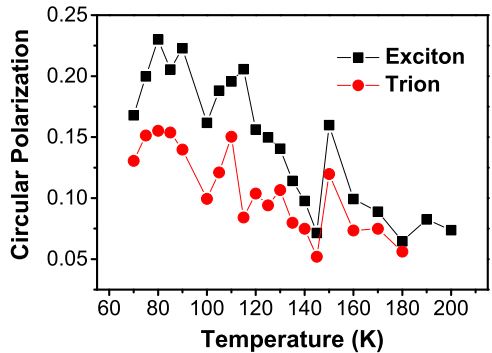}
  \caption{The temperature dependent circular polarization degree of both exciton and trion measured with pumping energy of 1.77eV.}
\end{figure}

\begin{figure}
\centering
  \includegraphics[width=8cm]{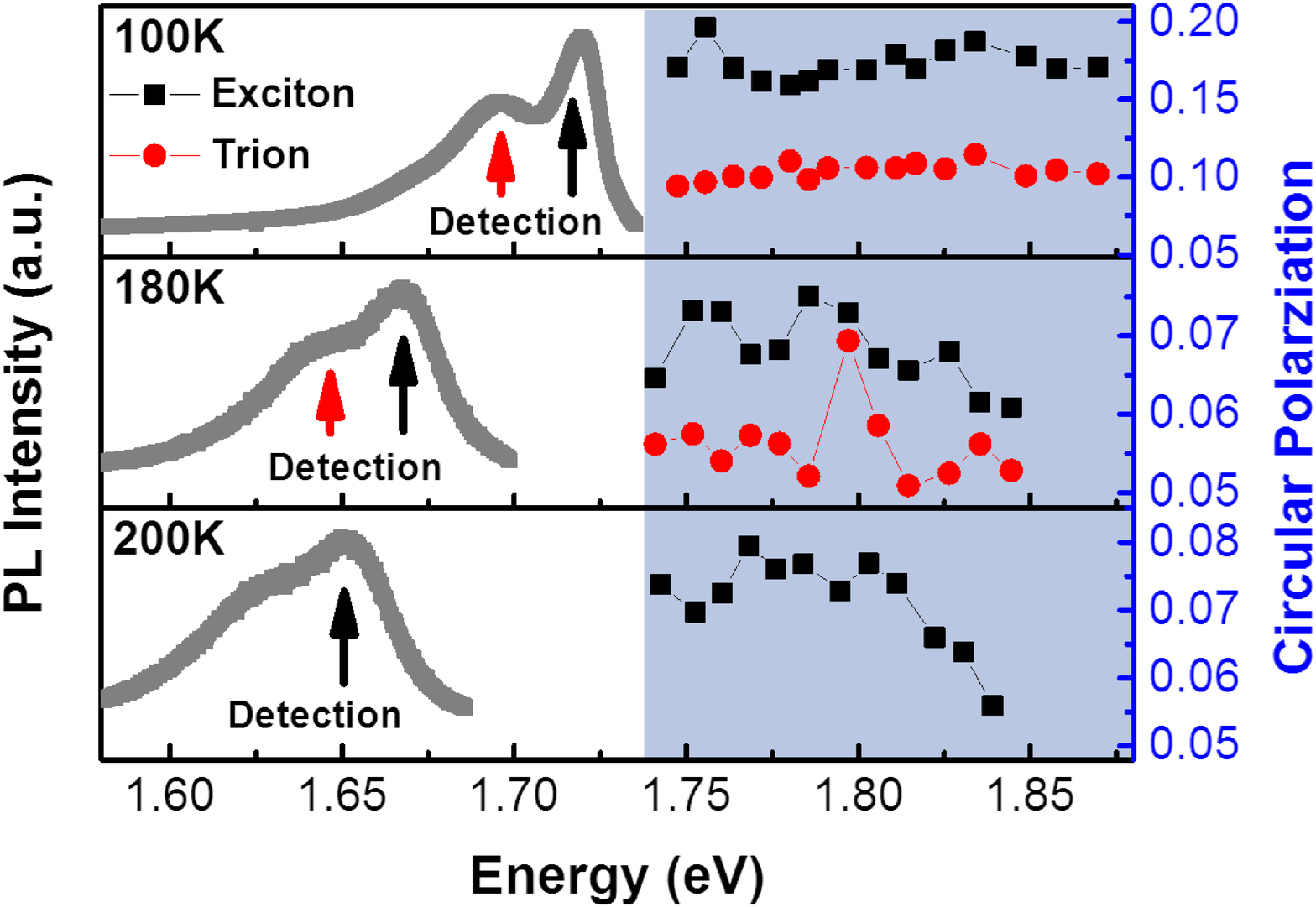}
  \caption{The circular polarization degree as a function of the excitation laser energy, shown in the blue-colored regime on the right. The grey lines on the left are the PL response at different temperatures, in order to show the energy separation between the A exciton and excitation laser.}
\end{figure}

The circular polarization of monolayer WSe$_2$ as a function of temperature (excited at 1.77eV) is investigated with the result shown in Fig. 3. As is clearly seen, the circular polarization degree decreases as temperature increases, which is consistent with previous works\cite{zeng2012valley,PhysRevLett.112.047401,PhysRevB.90.161302}. This can be understood as the momentum-dependent effective magnetic field $\Omega(P)$ is expected to increase with temperature, making spin precession and decoherence faster, thus reducing the exciton intervalley scattering time, leading to smaller PL circular polarization.

The excitation laser energy dependent PL circular polarization was investigated at different temperatures as well, with the results summarized in Fig. 4. Here the Fianium SC450-2 supercontinuum white light source has been employed together with a Semrock tunable bandpass filter to select the desired excitation laser energy. As can be seen in Fig. 4, when energy separation between A exciton and the excitation laser is smaller than 130meV, the measured circular polarization shows no obvious excitation laser energy dependence. This may be related to the complicated exciton fine structures.\cite{wang2014non} While when the energy separation between A exciton and the excitation laser is larger than 130meV, PL circular polarization is observed to decrease when increasing the energy separation. Similar trend has also been observed in previous works in MoS$_2$ \cite{kioseoglou2012valley,PhysRevLett.112.047401} and WSe$_2$ \cite{wang2014non}. This can also be understood based on the electron-hole exchange interaction model as the center-of-mass momentum of the A exciton is \cite{PhysRevB.89.205303}:

\begin{equation}
|P|=\sqrt{2m^*\varepsilon_{pump}}/\hbar,
\end{equation}
here $\varepsilon_{pump}$ is the excitation laser energy. Thus $|P|$ is propotional to the energy separation between A exciton and the excitation laser. From equations (2) and (3), it is obvious that the intervalley exciton precession frequency gets larger when $|P|$ increases. Hence, the valley(spin) relaxation time decreases with increasing excitation energy, \emph{i.e.}, lower PL circular polarization.

To explore the influence of exciton density on the exciton valley polarization, a picosecond laser (Fianium Ltd., model: SC450-2) with pulse duration of 40ps and a femtosecond Ti:Sapphire laser (Coherent Inc., model: Chameleon) with pulse duration of $\sim$150fs are used to increase exciton density in a wide order of magnitude. The estimation of the exciton density under excitation of two different laser systems is presented in Fig.6 in the supplementary material. The circular polarization of monolayer WSe$_2$ was first examined with optical pumping at 1.77eV with picosecond laser excitation by increasing the peak power intensity up to $10^8MW/cm^2$, which corresponds to a photo-generated exciton density of about $10^{14}cm^{-2}$. A typical result measured at 70K is presented in Fig. S2 in the supplementary material. Surprisingly, a non-monotonic dependence of circular polarization on exciton density is observed: one can see that the PL circular polarization first increases at relatively low exciton density and then falls at high exciton density with a maximum appearing at $\sim5.0\times10^{13}cm^{-2}$. However, the monolayer flakes can be occasionally burned under such high average laser power density. In order to have a repeatable and complete examination of exciton density denpendent valley polarization, we thus continued this study by using the femtosecond laser excitation to furthermore increase the exciton density while keeping the average excitation power low enough to avoid sample damaging. A more complete, non-monotonic dependence of valley polarization on exciton density can be well repeated as shown in Fig. 5. Our observation is different from the previous work\cite{PhysRevLett.112.047401}, in which only the decreased circular polarization with increasing excitation power has been observed, with the excitation laser pulse duration of 1.6ps and exciton density higher than $10^{13}cm^{-2}$, though a similar trend can be repeated in our experiment (shown in red dots in Fig. 5) under exciton density higher than $10^{13}cm^{-2}$. Meanwhile, TRPL measurements within the tuned excitation intensity range has also been taken in order to investigate the excitation intensity dependent exciton lifetime $\tau_0$ and valley lifetime $\tau_v$. It is found that the exciton lifetime $\tau_0$ does not depend on the excitation intensity within the examined excitation power intensity, while the excitation intensity dependent polarization relaxation time is hardly estimated due to the poor signal-to-noise ratio and the limitation of the temporal resolution (20ps) of the Hamamatsu streak camera system we used. However, based on equation (1), the excitation intensity dependent scattering rate can be deduced to exhibit a non-monotonic dependence on the exciton density.

\begin{figure}
\centering
  \includegraphics[width=8cm]{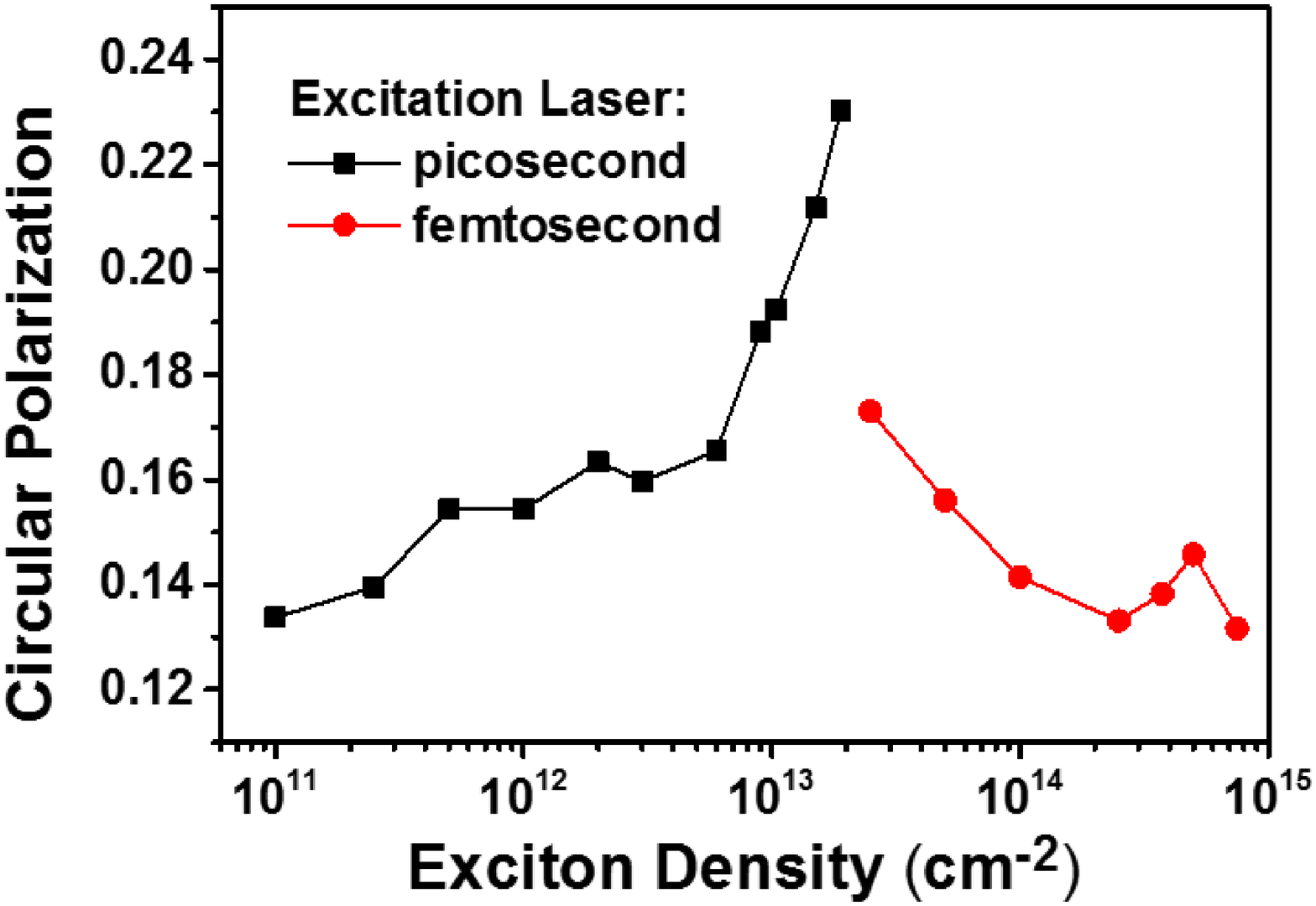}
  \caption{The PL circular polarization degree as a function of the injected exciton density, excited by picosecond and femtosecond laser, respectively.}
\end{figure}

In the past decade, there have been large amount of both theoretical and experimental studies for spin dynamics in III-V group semiconductors\cite{dyakonov2008spin,fabian2007semiconductor,wu2010spin}, in which DP mechanism has been revealed to dominate the spin relaxation process. The non-monotonic dependence of spin relaxation time on carrier density in GaAs quantum well structures has been previously observed\cite{PhysRevB.81.035213,PhysRevB.77.193307,han2011temperature}. Since the spin relaxation time is inversely proportional to the momentum scattering time $\tau_p^*$, \emph{i.e.}, motional narrowing, the spin relaxation time increases with carrier density in the non-degenerate (low carrier density) regime where $\tau_p^*$ decreases with increasing carrier density, whereas the inhomogeneous broadening $\langle\Omega^2(P)\rangle$ barely changes since carrier distribution function is described with the Boltzmann distribution\cite{wu2010spin,PhysRevB.79.125206,PhysRevB.89.205303,0256-307X-26-6-067201}. However, in degenerate (high carrier density) regime, the inhomogeneous broadening is greatly enhanced with electron density, leading to the decreased spin relaxation time with carrier density.

For the case of monolayer WSe$_2$ investigated in this work, in low excitation regime, the laser excited exciton density is low, \emph{i.e.}, the non-degenerate regime. Considering that the excited excitons within the non-degenerate regime obey the Boltzmann distribution, which is the same as that of electrons, we propose a similar relaxation process of valley pseudospin on exciton density as that of electron spin in GaAs quantum wells. The inhomogeneous broadening $\langle\Omega^2(P)\rangle$ originating from the long-range exchange interaction between electron and hole has weak dependence on the exciton density in non-degenerate regime, whereas the center-of-mass momentum scattering rate $1/\tau^*_p$ increases with increasing exciton density, hence the spin (valley) relaxation time $\tau_v$ increases with excitation density because of the motional narrowing, which is exactly what we observed experimentally as shown in Fig. 5. At higher excitation density, the large amount of excited excitons relax to the higher bands of the energy band as lower bands are filled. These higher-energy excitons, namely hot excitons, possessing larger center-of-mass momentum, precess with larger frequency due to the long-range e-h exchange interaction, leading to larger inhomogeneous broadening $\langle\Omega^2(P)\rangle$ and decreased valley(spin) relaxation time $\tau_v$. As a result, a peak for excitation density dependent valley polarization would appear at the crossover between the non-degenerate and degenerate exciton distribution regimes, as we observed in Fig. 5.

In summary, the circularly polarized PL response of monolayer WSe$_2$ has been systematically examined at different temperatures, excitation energies and intensities. The experimentally determined valley polarization and its relaxation provide evidence for the depolarization mechanism governed by the intervalley electron-hole exchange interaction. More importantly, we observe a non-monotonic dependence of valley polarization on the excitation power density, providing the experimental evidence for the non-monotonic dependence of exciton intervalley scattering rate on the excitation power density. The relaxation of valley pseudospin in monolayer TMDCs is thus proposed to exhibit the similar scenario as that of the DP-governed spin relaxation in conventional GaAs quantum well structures.

The valuable discussions with T. Yu and M. W. Wu are gratefully acknowledged. This work is supported by the National Natural Science Foundation of China (No. 11474276) and the National Basic Research Program of China (No. 2011CB922200).
\nocite{*}

\bibliography{apssamp}

\section{Supplementary Material}
\subsection{S1. Exciton density estimation excited by different laser systems}
To estimate the exciton density excited by different lasers, a rate equation was used:
\begin{equation}\label{S1}
  dN/dt=-N/\tau+Aexp(-(t-t_0)^2/(2\sigma^2 ))
\end{equation}

Where N is the exciton density and $\tau$ is the exciton lifetime. The second term in the above equation denotes the excited exciton increase described by a Gaussian function with parameters of the excitation laser, where ¦Ò is related to the laser pulse width and A describes the density of exciton excited by a laser pulse which is associated with the absorption coefficient of WSe2, average pulse energy and pumping photon energy. The exciton density is thus estimated under excitation of picosecond (with $\sim$ 40ps pulse width) and femtosecond (with $\sim$ 150fs pulse width) lasers, respectively, as presented in Fig 6, in which laser excited exciton density is plotted as a function of time with the same average laser power of 1¦ÌW for both laser systems.

\begin{figure}[h]
  \centering
  \includegraphics[width=8cm]{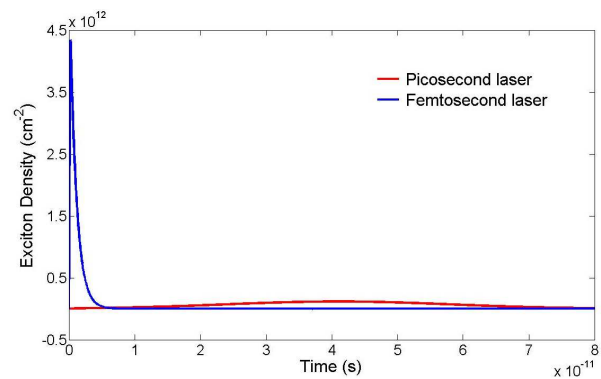}\\
  \caption{Calculated exciton density as a function of time under excitation by different laser systems. The blue and red lines correspond to the femtosecond and picosecond laser excitation, respectively.}
\end{figure}

\subsection{S2. Exciton density dependent photoluminescence circular polarization}

The circular polarization of monolayer WSe$_2$ was examined with optical pumping at 1.77eV with picosecond laser excitation first. A typical result of the exciton density dependent circular polarization degree by increasing the excitation power density of the picosecond laser, as described in the main text, is presented in Fig. 7. A clear peak of the circular polarization degree is seen at $\sim$5$\times10^{13}cm^{-2}$ with increasing exciton density measured at 70K. Note that the different circular polarization degree presented in Fig. 5 and Fig. 7 is due to the different flakes we used in the experiment. As mentioned in the main text, monolayer flakes tend to be damaged at high excitation laser intensity, thus the absolute circular polarization degrees slightly varies from piece to piece when examining different flakes.
\begin{figure}
  \centering
  \includegraphics[width=8cm]{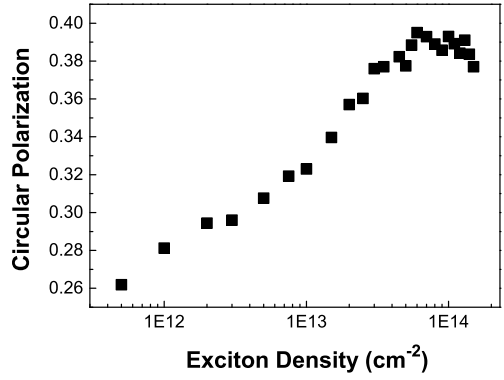}\\
  \caption{The PL circular polarization degree as a function of the injected exciton density, excited by the picosecond laser only at 70K and pumping energy of 1.77eV.}
\end{figure}
\end{document}